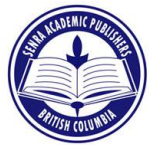

Canadian Journal of
pure&applied
sciences

# THE TWO-DIMENSIONAL DENSITY OF STATES IN NORMAL AND SUPERCONDUCTING COMPOUNDS

*P. Contreras[1], A. Devi[2], D. Uzcátegui[1] and E. Ochoa[1]
Physics Department, University of Los Andes, Mérida, 5001, Venezuela
Department of Physics, Government College Baroh, Kangra, Himachal Pradesh, India

## ABSTRACT

The present work represents a review for the numerical calculation of the density of states (DoS) for two-dimensional tight-binding models with first neighbors in its normal state and for two superconducting order parameters. One is a singlet scalar state and the other is a triplet vector state. At the beginning an emphasis is given to the general expressions commonly used to the calculation of the density of states as the number of partial and total number of states, the degrees of freedom and the *ab-initio* methods most commonly used to its calculation. Then, the transition happening to the DoS normal states by varying the Fermi energy and the hopping parameter is investigated. After that, the numerical calculation of the superconducting density of states using the zero-temperature scattering cross-section is performed for the two order parameters. Finally, the residual density of states depending on disorder and the scattering potential strength using the Larkin equation are calculated for the two order parameter symmetries different in nature.

## INTRODUCTION

The density of states (DoS) is a quantum mechanical concept derived from the total number of quantum states ($\Phi$) with an energy less than the value of $E$ in a microscopic system, where $\Phi$ provides the total number of states and increases in energy with the DoS defined as $\rho(E) = d\Phi(E)/dE$. Also, the DoS can be derived within a small interval of energies $\delta E$ as function of a reduced number of microscopic states ($\Omega$) with the DoS being the proportionality coefficient in the relationship $\Omega(E) = \rho(E)\delta E$, where $\Omega$ is constant in energy (Reif, 1965). On the other hand, the relation between the DoS and the degrees of freedom (*f*) states is the microscopic motion of a physical system that follows the Gibbs distribution (with a constant energy) and the behavior of some particles is quasi-classical and happens only for some degrees of freedom. However, for the rest of degrees of freedom the motion is quantized and those degrees of freedom can be written as function of a quantum number (*n*), where the energy is quantized as $E_n(q, p)$ (Landau and Lifshitz, 1969).

Additionally, the DoS can be function of external parameters such as those of extensive type defined in statistical thermodynamics (Reif, 1965) and in that case, different degrees of freedom can be included in a macroscopic system. Thus, another way to derive the DoS comes from comparing the Gibbs distribution and the microcanonical ensemble in Statistical Mechanics (Landau and Lifshitz, 1969). It is more complicated to understand the derivation that includes the Planck constant ($2\pi\hbar$) and the relation with a volume in a hyperspace with one Planck constant for each pair of the conjugate variables *q* and *p* in the phase space, where each microstate belongs to a $2fN$-dimensional "hypercube" with a length such as $2\pi\hbar$ and a volume such as $(2\pi\hbar)^{fN}$ (Landau and Lifshitz, 1969).

In general, the DoS it is a measure of how many microscopic states are available to a system in a particular range of values of the energy. If the ground state energy for a physical system with $N$ particles is given by $E_0 = fN\epsilon_0$, the energy difference from an excited state from the ground state is represented by a general expression linking several parameters. Thus, $E - E_0 = fN(\langle\epsilon\rangle - \epsilon_0)$, where the notation indicates that $\langle\epsilon\rangle$ is the mean quantum energy per particle, $\epsilon_0$ is the ground state energy of each particle, $N$ is the number of particles, $E$ and $E_0$ are the total energy and ground state energy of the $N$ particles, and *f* are the degrees of freedom. This mean that the total number of states $\Phi(E - E_0) \sim \Phi(\langle\epsilon - \epsilon_0\rangle)^N \sim (\langle\epsilon\rangle - \epsilon_0)^N$ is a big number even for only one kind (*f* = 1) of degrees of freedom.

*Corresponding author e-mail: pcontreras@ula.ve

In solid state physics, the DoS is expressed in terms of the system´s energy. As some authors point out (Mulhall and Moelter, 2014), each element of volume/area in the phase space of position $q$ and momenta $p$ is replaced by a weighting factor in an energy integral, which is easier to work with at the quantum level. We make use of the rationalized Planck units ($\hbar = k_B = c = 1$) to have a single unit of measurement because it conducts to the conceptual framework of the reduced phase space for the zero-temperature elastic scattering cross-section, which we have used previously in the following work (Contreras, *et al.*, 2023). A tale of the scattering lifetime and the mean free path. arXiv:2301.05322 [cond-mat.supr-con] DOI: 10.48550/arXiv.2301.05322).

Thus, a DoS equation with equal number of spin up and down particles can be written as: 1) the sum of infinite delta functions; 2) as the derivative of the total number of states; 3) as a proportionality coefficient of the partial number of states, i.e.

$$N(\omega) = {}^2/_V \sum_i^\infty \delta(\omega - \omega_i) = {}^{d\,\Phi}/_{d\,\omega} \approx {}^\Omega/_{\delta\,\omega}$$

The last two expressions relate the DoS and the number of total or partial states. An instructive interpretation of the DoS from a geometrical perspective is given in (Mulhall and Moelter, 2014) where the DoS is defined as the slope between the number of macroscopically allowed quantum partial number of states ($\Omega$) and an infinitesimal energy interval from $\omega$ to $\omega + \delta\omega$ in a two-dimensional space with variables ($\omega, \Omega$) (Mulhall and Moelter, 2014).

The sum inside the delta function in momentum space is adequate for the normal state DoS since disorder only changes the DoS value by a constant quantity. But in the other cases such as unconventional superconductors, it is easy to replace the sum by a weighting factor into an energy integral, which is easier to deal with. Summarizing, the DoS is extensively used in applications to Statistical Mechanics, Solid State Physics, and Quantum Chemistry. *Ab-initio* routines include the calculation of the DoS, and more important is that the DoS can be calculated not only for systems with $N$ particles but also for: 1) single molecules at the Hartree-Fock HF/6-311G* level (Contreras *et al.*, 2021) using the TDoS formalism (Lu and Chen, 2012); 2) dimer or trimer molecular systems with lack of inversion symmetry at the UDFT/B3LYP level (Burgos *et al.*, 2017), or in one and two-dimensional monolayers such as nanowires and nanoflakes, where it is clearly observed from several DoS calculations and their visualizations show that the materials symmetrical for its up and down channels are nonmagnetic and asymmetrical materials are magnetic in nature (Devi *et al.*, 2019, 2021a, 2021b, 2022).

The structure of this work is as follows: In the following section, some details of the computational approach are outlined. In the third section, a detailed DoS calculation of the normal state with a tight binding model is performed. In the fourth section, the calculation of the superconducting DoS is performed for singlet and a triplet order parameters (OP). In the final section, the behavior of the residual density of states is addressed for both models using the formalism following the Larkin equation (Larkin, 1965).

**Computational details for the density of states with sums and Fermi averages**

For the numerical calculation of the DoS in the normal state, we make use an approximation of the Delta function using a 2D sum for momentum dependency with a $\delta$-function approximated by

$$N(E) \cong \sum_{kx,ky} \delta(E - \xi_{i,k}) = \frac{1}{\pi} \sum_{kx,ky} \frac{n}{\left(E - \xi_{i,k}\right)^2 + n^2} \quad (1)$$

where $\Sigma_{kx,ky}\delta(E - \xi_{i,k})$ is approximated by a Lorentzian 2D function. For the calculation of equation (1), it has been used a mesh of ($N \times N$) $k$-points with $N = \pm\,400$ points. The other parameters in equation (1) are $n = 0.005$ that gives a well-defined delta function, and a dispersion TB law with first neighbors where two terms are responsible for the behavior of the DoS, i.e.

$$\xi_{i,k}(k_x, k_y) = \varepsilon_F + \xi_{hop}(k_x, k_y) \quad (2)$$

where the $\mathbf{k}$ dependence is carried in the hoping term $\xi_{hop}(k_x, k_y)$. Using the first neighbors' coefficient $t$, this function is given by

$$\xi_{hop}(k_x, k_y) = 2t \left[ \cos\left(^{k_x \pi}/_N\right) + \cos\left(^{k_y \pi}/_N\right) \right]$$

On the other hand, for a superconductor with nonmagnetic impurity scattering it is used the equation which is derived from the Green function formalism (also known as T matrix formalism) (Mineev and Samokhin, 1999; Hussey, 2002) $N(\widetilde{\omega}) = N_F \Re[g(\widetilde{\omega})]$ where $\Re$ means the real part of the function $g(\widetilde{\omega})$. The Fermi level DoS is $N_F$ and the function containing the impurity effects is

$$g(\widetilde{\omega}) = \langle \frac{\widetilde{\omega}}{\sqrt{\widetilde{\omega}^2 - \Delta_0^2(k_x, k_y)}} \rangle_{FS}$$

and has a zero superconducting energy gap parameter dependence. The function $g(\widetilde{\omega})$ has the Fermi average $<...>_{FS}$. This part requires a calculation that implies uncommon numerical routines to find from the zero self-consistent elastic scattering cross-section, the real and imaginary parts. The study of the zero-temperature elastic scattering cross-section was firstly proposed in (Pethick and Pines, 1986) and used with a specific disorder parametrization, i.e. the inverse dimensionless strength $c$ and the impurity density $\Gamma^+$ in (Schachinger and Carbotte, 2003) and references therein for isotropic Fermi surfaces.

Additionally, extended studies of the zero elastic scattering cross-section were recently performed in (Contreras and Moreno, 2019) to calculate $\overline{\omega}$ using two different numerical self-consistent routines for isotropic FS and a linear nodes OP with different $c$ and $\Gamma^+$ in order to establish differences in numerical routines. The work by Contreras and Osorio (2021) was performed for a linear nodes HTSC model using a tight binding parametrization for three different collisional regimes. For the for the Miyake Narikiyo quasi-nodes triplet OP (Miyake and Narikiyo, 1999), the tight-binding calculation of $\overline{\omega}$ was performed for ten values of the inverse strength parameter $c$ showing that the cross-section is mostly in the unitary limit and few times in the intermedium limit (Contreras *et al.*, 2022a).

In (Contreras *et al.*, 2022b), the calculation was performed as function of the Fermi energy and it was distinguished the point nodes model from the quasi-nodal original model in the elastic scattering cross-section. In (Contreras *et al.*, 2022c), the dependence on the zero temperature $\Delta_0$ was modeled self-consistently for a triplet OP finding that the imaginary elastic scattering cross-section is always positive and fits well in the unitary limit. Finally, the quasi-nodal model was contrasted with the linear OP behavior by fixing the Fermi energy and the zero-temperature superconducting gap in other to see the interplay between different kind of quasiparticles (Kaganov and Lifshits, 1989) in the elastic scattering cross-section (Contreras *et al.*, 2022d, 2022e).

If we are dealing with more than one Fermi surface sheet, the DoS is calculated according to equations such as

$$\frac{N(\omega)}{N_F} = p^\gamma N^\gamma \left(\frac{\omega}{\Delta_0^\gamma}\right) + p^{\alpha,\beta} N^{\alpha,\beta} \left(\frac{\omega}{\Delta_0^{\alpha,\beta}}\right)$$

which is suitable for strontium ruthenate in a non-self-consistent way using a TB parametrization aiming at fitting experimental low-temperature data such as ultrasound attenuation in the superconducting state $\alpha(T)$ (Lupien *et al.*, 2001; Contreras *et al.*, 2004), the electronic superconducting thermal conductivity $\kappa(T)$ (Tanatar *et al.*, 2001; Contreras, 2011) and the electronic superconducting specific heat $C(T)$ (Nishizaki *et al.*, 1998; Contreras *et al.*, 2014) with relatively clean samples. Details of the original use of $p^\gamma$ and $p^{\alpha\beta}$ are found for the $Sr_2RuO_4$ normal state viscosity calculation in (Walker *et al.*, 2001). The other work of relevance for experimental fittings in strontium ruthenate can be found in (Nomura, 2005; Taniguchi *et al.*, 2015; Wu and Joynt, 2001; Zhitomirsky and Rice, 2001).

The equation used to calculate the DoS in dirty superconductors when taking into account the reduced phase space is the following one (Devi *et al.*, 2021b):

$$\frac{N(\widetilde{\omega})}{N_F} = \langle \frac{\mathfrak{R}(\widetilde{\omega})}{\sqrt{2\,\rho_k}} \sqrt{1 + \frac{a_k}{\rho_k}} \rangle_{FS} + \langle \frac{\mathfrak{I}(\widetilde{\omega})}{\sqrt{2\,\rho_k}} \sqrt{1 - \frac{a_k}{\rho_k}} \rangle_{FS} \quad (3)$$

where $\mathfrak{R}(\widetilde{\omega})$ and $\mathfrak{I}(\widetilde{\omega})$ are the coordinates in the reduced phase space, and the other symbols are $a_k = \mathfrak{R}(\widetilde{\omega})^2 - \mathfrak{I}(\widetilde{\omega})^2 - \Delta_k^2$, $b = 2\mathfrak{R}(\widetilde{\omega})\mathfrak{I}(\widetilde{\omega})$, and $\rho_k = (a_k^2 + b^2)^{1/2}$. Equation (3) is a very suitable because it is directed related to the reduced phase space.

The tight-binding Fermi averages replacing the sum "$\Sigma_{kx,ky}(\dots)$" are performed using a weight in energy instead of the sum, i.e.

$$\sum_k^\infty (\dots) = \frac{1}{4\,\pi^2} \int \frac{d\,S_F}{|\overrightarrow{v_k}|} \int d\,E\,(\dots) = \langle \cdots \rangle_{FS},$$

where $E$ is the energy of the normal state, the Fermi velocity is given by the gradient $\overrightarrow{v_k} = \mathrm{grad}(E)$, the surface $k$-space element is expressed as $dS_F = (dk_x^2 + dk_y^2)^{1/2}$, and the normal state density of states at the Fermi level is calculated by Ashcroft and Mermin (1976) as follows:

$$N_F = \frac{1}{4\,\pi^2} \int \frac{d\,S_F}{|\overrightarrow{v_k}|}$$

## Normal state DoS and its evolution according to the Fermi energy

In this section, we demonstrate how the evolution of the density of states (DoS) can be numerically modeled by varying the Fermi energy parameter in equation (2). Using the results shown in Figure 1, it is found that the implicit Fermi surface with $\xi_{i,k}(k_x, k_y) = 0$ evolves from having a behavior with a mesh centered at (0, 0) coordinates when the Fermi energy has a negative sign and the hoping coefficient has a positive sign, to a different behavior when both the Fermi energy and the hoping parameter have positive values, and the Fermi surface for this case is centered in four pockets at the $\pm$ (N, N) corners. The values used for the model are listed in Table 1. The implicit Fermi surface evolution in the $N \times N$ mesh is sketched in Figure 1, where two well defined behaviors are seen.

For the values of the parameters listed in Table 1, two of the four implicit plots are centered at zero point (when the hoping parameter value is $t = 0.20$ meV, i.e. the green and red implicit plots). Meanwhile, the other two are centered at the corners of the square (when the hoping parameter value is $t = + 0.40$ meV, i.e. the yellow and blue implicit plots). In addition, in this section, the normal density of states is calculated using equation (1) with the same parameters listed in Table 1. The results are shown in Figure 2.

Table 1. The evolution of the implicit Fermi surface for different sets of TB parameters.

| Fermi energy $\varepsilon_F$ | Hoping parameter $t$ | Centered at $N \times N$ mesh points | $\xi_{i,k}(k_x, k_y) = 0$ |
|---|---|---|---|
| − 0.40 meV | + 0.20 meV | ± (400, 400) | Red color |
| − 0.04 meV | + 0.20 meV | ± (400, 400) | Green color |
| + 0.04 meV | + 0.40 meV | (0, 0) | Blue color |
| + 0.40 meV | + 0.40 meV | (0, 0) | Yellow color |

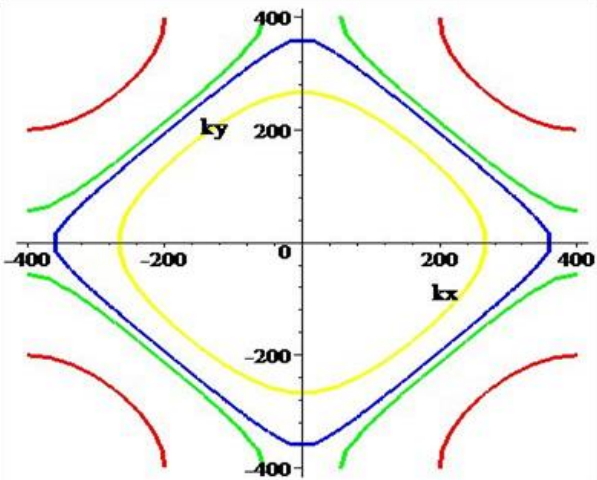

Fig. 1. The evolution of $\xi_{i,k}(k_x, k_y) = 0$ in a $400 \times 400$ points mesh. The TB values for each color are listed in Table 1.

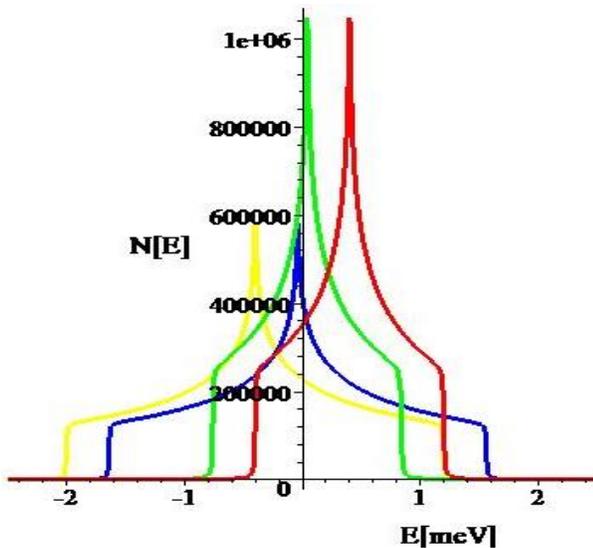

Fig. 2. The density of states $N(E)$ for 4 values of $E_F$ in a $400 \times 400$ points mesh. The TB values for each color are from Table 1 and Figure 1.

As it can be seen in Figure 2, the density of states with Fermi energy values close to zero (blue and green colors) have almost an electron-hole symmetry behavior. Also, those where the two tight binding coefficients have similar order of magnitude are less symmetric (yellow and red colors). If the hoping parameter is smaller as happen for the red and green cases, both centered at the corners, the DoS have more available quantum states than when the Fermi surface is centered at zero points. This is important and shows how the quantum behavior of those ceramics having tight-binding parametrization (the red and green cases) (for instance, some HTSC in its normal state) represent a more difficult quantum interpretation compared with the blue and yellow cases that mostly represent metallic alloys.

The drop to zero of the normal DoS can be understood in terms of a number of partial available quantum states constant as it has been explained in the introduction. Therefore, the DoS being a coefficient of the partial number of states $\Omega$ (see the introduction of this work) becomes negligible and probably other degrees of freedom start to play a more important role at those energies where $N(E)$ drops to a zero value, meaning a constant number of partial states $\Omega$.

**Impurity superconducting DoS and its evolution as function of the scattering strength and disorder**

In the superconducting state, we can compare two OP models. The 2D TB line nodes used to model the strontium doped lanthanum copper oxide superconductor with a $T_c \approx 44.35$ K for a polycrystalline sample (Bednorz and Müller, 1986; Kastner et al., 1998; Xiao et al., 1987) modeled with the parameters $t = 0.2$ meV, $\varepsilon_F = −0.4$ meV and $\Delta_0 = 33.9$ meV (Yoshida et al., 2012) shadowed gray in the 1st line of Table 1. The linear nodes OP has even parity $i$ that belongs to the irreducible representation B$_{1g}$ of the D$_{4h}$ point symmetry group (Scalapino, 1995; Tsuei and Kirtley, 2000).

The triplet case is represented in this section for the 2D $\gamma$-sheet Miyake Narikiyo quasi-point nodes model (Larkin, 1965) for strontium ruthenate, and $T_c \approx 1.5$ K for a bulk clean sample (Maeno et al., 1994; Rice and Sigrist, 1995) with the following values: $t = 0.4$ meV, $\varepsilon_F = −0.4$ meV, and $\Delta_0 = 1.0$ meV shadowed gray in the 4th line of Table 1. In this case, the OP has odd parity $i$ that belongs to the irreducible representation E$_{2u}$ of the D$_{4h}$ point symmetry group with GL coefficients (1, $i$) and 2D basis (sin($k_x$),

sin($k_y$)) (Miyake and Narikiyo, 1999; Walker and Contreras, 2002; Sigrist, 2002; Contreras *et al.*, 2016). We would like to point out the intriguing 2D electronic nature of this material as is pointed out in (Maeno *et al.*, 1997).

In addition, to take into account the nonmagnetic disorder effects in both order parameter models inside the superconducting DoS and the residual DoS, we use:

- The Born limit given by $l\,k_F \gg 1$ or $l\,a^{-1} \gg 1$, where $l$ is the mean free path, $a$ is the lattice parameter, and $k_F$ is the Fermi momentum.
- The intermediate scattering regime with $l\,k_F \sim l\,a^{-1} > 1$.
- The unitary limit where holds that $l\,k_F \sim l\,a^{-1} \sim 1$.

We use the parameter $c$ which is inverse to the scattering strength $U_0$ to describe the dispersion limits for both OP numerical models. During the 1970s and the 1980s, the formalism and some phenomenology of the physics for nonmagnetic impurity scattering in normal metals and alloys were described in (Ziman, 1979) from a work firstly proposed by Edwards (1958). In monography (Lifshits *et al.*, 1988), it was noticed that in studying metallic alloys, there are singularities in the DoS for disordered systems such as those involving nonmagnetic impurities, or stoichiometric nonmagnetic atomic potentials $U_0$, they pointed out that keeping a 1st power in impurity concentration suffixes the calculation. We consider also instructive to mention that the ARPES technique is related to a fundamental equation "*the Fermi Golden Rule*". A robust introduction to ARPES can be found in (Palczewski, 2010).

## Numerical results for the self-consistent DoS with nonmagnetic disorder

Figure 3 shows the density of states (DoS) calculated numerically using equation (3) that includes the zero elastic scattering cross-section reduced phase space. The resulting plot is the density of states $N(\overline{\omega})/N_F$ as a function of the normalized frequency $\omega/\Delta_0$. In the clean limit (without impurities), the parameter for the concentration of disorder $\Gamma^+$ is normalized by the zero-energy gap and denoted as $\zeta = \Gamma^+/\Delta_0$. It is observed that, there are no dressed fermionic quasiparticles if there is no disorder. Thus, the number of occupied states starts to grow linearly, since it represents a nodal line OP (Scalapino, 1995) until an energy value equal to that of the zero superconducting gap is reached. At $\omega = \Delta_0$ there is a drastic change in slope that resembles on the right side the shape of a BCS superconductor (Bardeen *et al.*, 1957), where the DoS represents a singularity for $\omega = \Delta_0$ and is equal to zero below the gap, the other type of suppression in superconducting states (when there are magnetic impurities) is explained following a physical kinetic analysis in (Ambegaokar and Griffin, 1965).

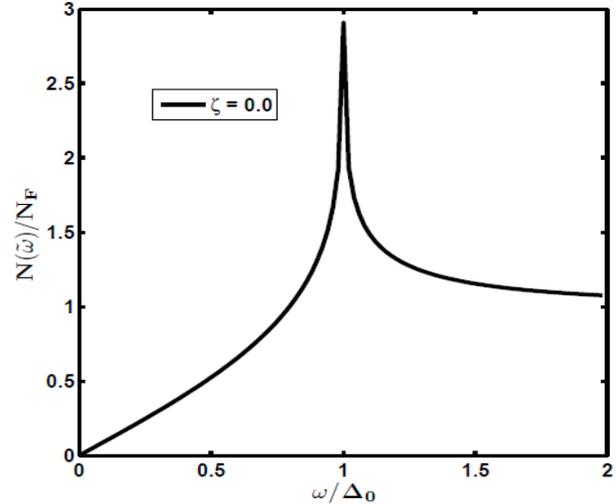

Fig. 3. The superconducting density of states (DoS) for a tight binding line nodes OP if $\zeta = 0$. There aren't residual states at zero energy.

Figure 4 shows five curves calculated for the values of disorder with $\zeta = 0.001, 0.005, 0.010, 0.015$, and $0.020$ in the unitary limit when $c = 0$. However, at zero frequency, it is observed that there is a residual density of states for all 5 values of $\zeta$, which is a consequence of the presence of nonmagnetic impurities in the reduced phase space for the line nodes model, which behavior is modified by the inverse scattering lifetime. It is noticed that the higher concentration of impurities, the number of states at the frequency value of $\omega = 0.0$ meV is also higher. The residual normalized DoS is bigger for the thinner line ($\zeta = 0.020$) compared to other four curves, the thicker line ($\zeta = 0.001$) has still a considerable number of residual DoS.

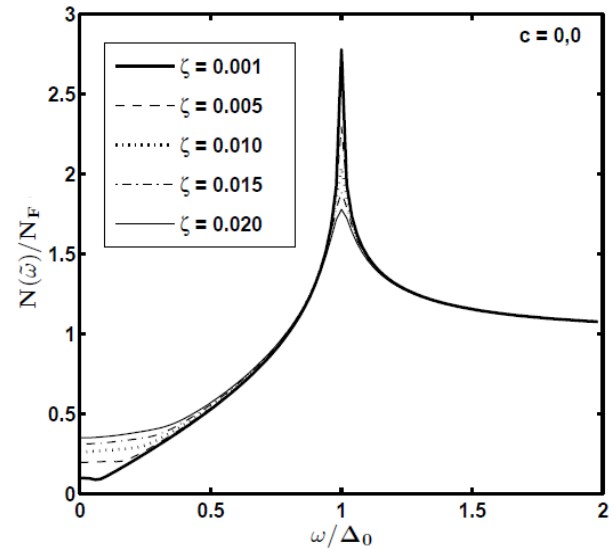

Fig. 4. The superconducting DoS for the line nodes OP. The residual DoS shows significant dressed quantum levels at zero energy due to the strong scattering potential.

Figure 5 shows the superconducting DoS for a weaker scattering potential with $c = 0.4$ and the disorder parameter $\zeta = 0.001, 0.005, 0.010, 0.015,$ and $0.020$. The strength when $c = 0.4$ was established numerically from the analysis of the zero-temperature elastic scattering cross-section as the Born limit for a linear OP (Contreras and Osorio, 2021). However, in this case the residual $N(0)/N_F$ weakly increases as $\zeta$ increases and it is noticeably smaller compared to the $N(0)/N_F$ values shown in Figure 2 that corresponds to the unitary limit, where there is a strong elastic scattering, suggesting that a strong nonmagnetic dispersion potential produces more occupied states than weaker scattering potentials at zero energy. This can be defined as the signature for the unitary state in the residual DoS analysis compare with the nonlocal minimum observed in the analysis of the imaginary part of the elastic cross-section $\Im[\widetilde{\varpi}](\omega + i0^+)$ (Contreras and Osorio, 2021).

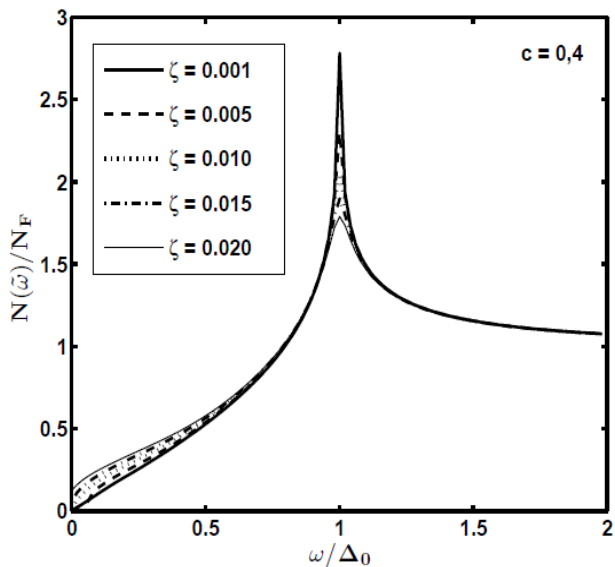

Fig. 5. The DoS for the nodal line OP in strontium doped lanthanum compound with 5 disorder values. The residual density is small when compared to the results shown in Figure 4. There are only a few quantum states available in the hydrodynamic limit.

The next calculation is performed for the $\gamma$ sheet of the triplet superconductor strontium ruthenate with an OP that belongs to the irrep $E_{2u}$ of the $D_{4h}$ point group. Figure 6 shows six curves calculated for different values of the normalized disorder $\zeta = \Gamma^+/\Delta_0$. The simulation was performed for $\zeta = 0.00, 0.01, 0.02, 0.03, 0.10, 0.20$ in the unitary limit when the inverse elastic scattering parameter $c = 0$, according to the analysis (Miyake and Narikiyo, 1999; Maeno et al., 1994; Rice and Sigrist, 1995; Walker and Contreras, 2002; Sigrist, 2002; Contreras et al., 2016; Maeno et al., 1997). Triplet OP are suitable to analyze in this limit due to strong nonmagnetic strontium potential. A contrast concerning the previous case shown in Figures

3, 4, and 5 is that the dimensionless disorder parameter for the triplet case is one order of magnitude bigger than the dimensionless singlet OP.

This is partially explained because the $\gamma$-sheet uses an experimental zero gap value of $\Delta_0 = 1.0$ meV that is an order of magnitude smaller that the lines' nodes OP with a zero gap $\Delta_0 = 33.9$ meV and therefore, the triplet case has a smaller reduced phase space for scattering events. Additionally, in the triplet compound, Sr atoms are located in the lattice with an additional nonmagnetic impurity level in the energy zone. Thus, Sr atoms are part of the $D_{4h}$ tetragonal structure and also are the scattering centers that explains the stronger pair breaking mechanism and the additional impurity level.

Figure 6 shows that in the absence of impurity levels (black line where $\zeta = 0$), we do not observe nonzero values for the density $N(\overline{\varpi})/N_F$ as happens for BCS superconductors (Bardeen et al., 1957). It means that when doing calculations that involved triplet states and scattering is excluded, there is no pair breaking effects and bosonic quasiparticles dominate the behavior below the transition temperature. It occurs numerically below 0.83 meV in this calculation. Generally speaking, this value depends on the choice of the TB parameters, i.e. how close will be the Fermi surface to the zero gap value of $\Delta_0$ in the MN model. We have used parameters from the $4^{th}$ column in Table 1 (shadowed gray) to calculate the DoS.

For the parameter value of $\zeta = 0.01$, we observe still an intermedia well-formed BCS gap from frequencies in the interval $(0.4, 0.83)$ meV and with a small quantity of quantum states at low frequencies due to strong scattering that happens in the unitary regime when the reduced phase space is activated with a nonzero imaginary part of the cross-section, and the mean free path $l$ is comparable to the magnitude of the inverse Fermi length $|k_F|^{-1}$, or to the value of the lattice parameter $a$. To illustrate what happens numerically we show in the insert on the upper left side of Figure 6, the reduced phase space calculation, and it is observed the following: for the case of $\zeta = 0.01$, the imaginary part of the cross-section dies inside of the superconducting phase. Therefore, it does not become a normal metal and could be a signature of an antiferromagnetic state as happens to the antiferromagnetic insulator LaCuO (Kastner et al., 1998).

For the impurity value of $\zeta = 0.05$, it agrees with an inhomogeneous phase that is the Miyake-Narikiyo tiny gap inside of which there are not fermionic quantum levels. Henceforth, the tiny Miyake-Narikiyo tiny gap predicted and used to explain microscopically the behavior of triplet pairing superconductors such as strontium ruthenate (Miyake and Narikiyo, 1999), is shown in Figure 6 for an impurity value of $\zeta = 0.05$ as was also observed in (Contreras *et al.*, 2022a, 2022b, 2022c) using the imaginary part analysis of the scattering cross-section $\Im[\tilde{\omega}](\omega + i0^+)$. The DoS calculation also agrees with the one by Contreras *et al.* (2022a) in the sense that only the unitary limit persists in Sr$_2$RuO$_4$. It is noticed that the higher concentration of impurity levels given by the values of $\zeta$ = 0.10, 0.15, 0.20, the larger number of occupied quantum states exists at both the low and high frequencies. This is a consequence of having an increasing reduced phase space and therefore more possibilities for scattering events. However, it is still small compared with the HTSC nodal lines' OP case (Contreras *et al*., 2022d).

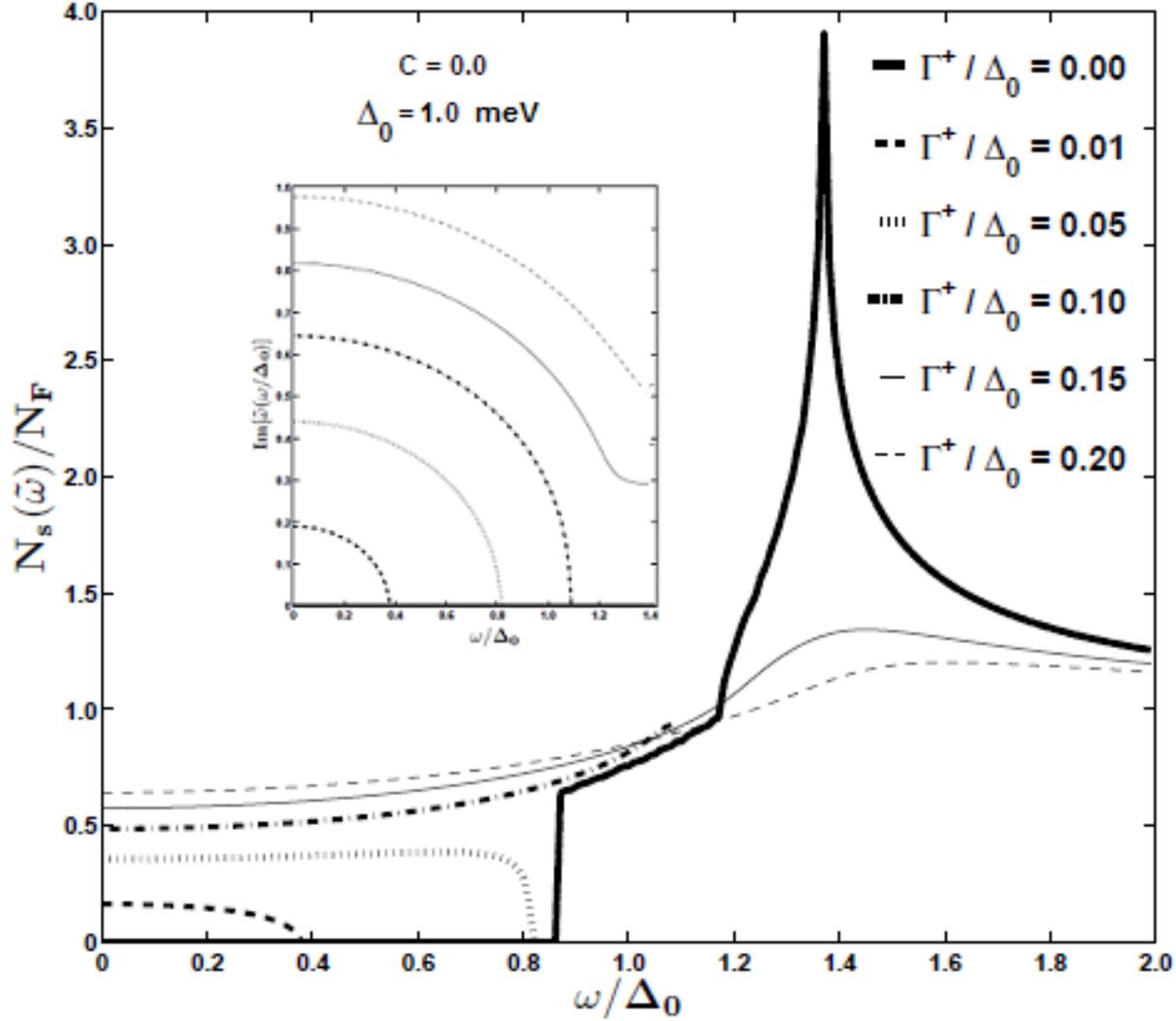

Fig. 6. The DoS for the quasi-point nodes with 6 values of nonmagnetic impurity doping including the clean case. The residual $N(0)$ is found for 5 nonzero $\zeta$-values. The tiny MN gap is found for $\zeta = 0.05$.

Table 2. The $C_0$ parameter and the residual DoS. It is summarized for both irrep, $B_{1g}$ and $E_{2u}$ and the scattering limits.

| Residual density of states RDoS TB formalism | Expressions for $C_0$ in the Born and intermedia limits | Expressions for $C_0$ in the unitary limit | Residual DoS for the Born and intermedia regimes | Residual DoS for the unitary limit |
|---|---|---|---|---|
| Singlet parity linear nodes OP | $C_0 = \pi\eta_c \, \langle \frac{\sqrt{C_0^2 + \phi_k^2}}{C_0} \rangle_{FS}^{-1}$ | $C_0 = \pi\eta_c \, \langle \frac{\sqrt{C_0^2 + \phi_k^2}}{C_0} \rangle_{FS}$ | $\frac{N(0)}{N_F} = \frac{C_0}{\pi\eta_c}$ | $\frac{N(0)}{N_F} = (\frac{C_0}{\pi\eta_c})^{-1}$ |
| Triplet parity quasi-point nodes OP | $C_0 = \pi\eta_c \, \langle \frac{\sqrt{C_0^2 + |d_k^2|}}{C_0} \rangle_{FS}^{-1}$ | $C_0 = \pi\eta_c \, \langle \frac{\sqrt{C_0^2 + |d_k^2|}}{C_0} \rangle_{FS}$ | $\frac{N(0)}{N_F} = \frac{C_0}{\pi\eta_c}$ | $\frac{N(0)}{N_F} = (\frac{C_0}{\pi\eta_c})^{-1}$ |

Therefore, for the disorder parameter values of $\zeta$ = 0.10, 0.15, 0.20, there are normal state DoS levels available, and the peak at $\omega$ = 1.4 meV considerable reduces with a tendency where $N(\omega) \sim N_F$ above $T_c$. Therefore, in the superconducting triplet model, we observe two phases, one tiny phase (the MN gap) without electronic levels (BCS type) and another with normal-state dressed quantum levels, contrasting with the strontium substitute lanthanum cuprate calculation, where dressed fermionic levels are found.

**Numerical results for the residual density of states and the pair breaking**

The residual equation for the density of states is theoretically obtained by setting up the real frequency as an imaginary number, i.e. $\omega = i\alpha$ in equation (4). Thus, we have $\widetilde{\omega} = \omega + i\alpha$ with $\alpha$ being a new disorder parameter ($0 \leq \alpha \leq 1$). This does not require a self-consistent routine but it needs a fixed-point numerical calculation. In such a case, we get the following general equation for the residual DoS ($N(0)/N_F$) with a new dimensionless disorder parameter $C_0 = \alpha/\Delta_0$:

$$N(0) = N_F \left\langle \frac{C_0}{\sqrt{C_0^2 + \Delta_0^2(k_x, k_y)}} \right\rangle_{FS} \qquad (4)$$

where equation (4) depends on the symmetry of the OP, the elastic scattering regime of the imaginary part of the zero-temperature elastic scattering cross-section, i.e. the unitary, intermedia, and Born cases; and finally also depends on the Fermi surface average, so we can control the residual DoS the same way as we did for the zero-temperature elastic scattering cross-section and the superconducting DoS.

There is the functional dependence $T_c/T_{c0} = f(N(0)/N_F)$, where $T_{c0}$ indicates the transition temperature without disorder and $T_c$ is the transition temperature including disorder in the Larkin equation (Larkin, 1965) for suppression of impurity states in the case of nonmagnetic disorder when the critical temperature $T_c$ decreases as a function of the pair breaking parameter $\eta_c$. So, we have

$$\ln \frac{T_c}{T_{c0}} = \psi\left(\frac{1}{2}\right) - \psi\left(\frac{1}{2} + \eta_c\right) = \psi'\left(\frac{1}{2}\right)\frac{\Gamma^+}{2\pi T_c} \qquad (5)$$

In equation (5), the superconducting pair breaking parameter is defined as $\eta_c = -(4\pi/1.764) \times \ln(T_c/T_{c0})$, where $T_{c0}$ is the transition temperature for a clean superconductor ($\alpha$ = 0), $T_c$ represents the transition temperature for dirty superconductors, i.e. $\alpha \neq 0$, $\psi(x)$ is the digamma function, and $\psi'(x)$ is the derivative of the digamma function. Table 2 lists the expressions that can be obtained if the tight binding approximation is accounted for and they do not differentiate from the isotropic case with angular dependence of the Fermi surface.

The analytical expressions for the calculation of the relationship $T_c/T_{c0} = f(N(0)/N_F)$ for the cases plotted in this section can be obtained after some long algebraic manipulations using equations (7) and (8). The difference from the previous work by Sun and Maki (1995) and Momono and Ido (1996) is that the Fermi surface average depends on more parameters $\langle ... \rangle_{FS}$, the basis function $\phi_k$ in the case of a scalar line nodes OP, and the complex triplet vector OP $d_k$ are the same used for the density of states calculation in the previous paragraph, the tight binding parameters are those shadowed gray in Table 1. A discussion with second and third harmonics for the triple OP is given in Miyake-Narikiyo original work (Miyake and Narikiyo, 1999).

Theoretically it is known that nonmagnetic impurities destroy superconductivity in unconventional superconductors (Larkin, 1965; Sun and Maki, 1995) and reduce the value of the transition temperature $T_{c0}$. The residual DoS changes as a function of $T_c$, as it is shown accordingly to the data listed in Table 2. Noticeable in this work is that the parameter $C_0$ depends on the Fermi surface averages. In order to numerically evaluate the polynomic expressions involved, it is more convenient to simplify the analysis to the three cases: The Born, intermedia and unitary scattering regimes (see in Table 2 for a summary of the equations involved).

The results of the numerical calculation using equations taken from Table 1 are shown in Figure 7 for the case of linear nodes OP. At $T_c/T_{c0}$ = 0 the largest residual DoS value for both the cases is obtained (for unitary and Born limits). On the other hand, as the value of $N(0)/N_F$ increases, the ratio $T_c/T_{c0}$ can rapidly reach a zero value, faster for a weak scattering Born nonmagnetic potential limit than for the unitary case. In addition, we see in Figure 5 that the unitary limit presents a curve that always has the same sign in slope; meanwhile the Born limit changes it signs and even has a linear behavior dependence for $N(0) \sim 0.5N_F$.

Figure 8 compares the unitary limit of the OP used with measurements taken from specific heat in the compound strontium doped lanthanum ceramic for different experimental values of strontium doped obtained experimentally (Momono and Ido, 1996). Strontium doping has been extensively studied in the cuprate La$_{2-x}$Sr$_x$CuO$_4$ for smaller orders of strontium concentration (Yoshida *et al.*, 2012; Momono and Ido, 1996). The color points with the impurity atoms are from values of specific electronic heat capacity $C(T)$ in the superconducting state, where in Figure 6 the gray color corresponds to $x = 0.10$, the green color to $x = 0.18$, the blue color to $x = 0.20$, and the red color to $x = 0.22$ (Momono and Ido, 1996). We see a tendency for the experimental red points with $x = 0.22$ corresponding to our reduced value of $\zeta = 0.02$ in correspondence with both the unitary theoretical residual density of states and the self-consistent unitary case of the previous section (Fig. 4).

In Figure 9, the value of $T_c/T_{c0}$ approaches zero slowly if the fixed-point calculation is done for the unitary limit, as in the case of line nodes. Meanwhile, the unitary triplet OP presents a curve that always has the same shape and slope; the intermediate limit changes its slope weaker, slightly contrasting with the OP nodal line situation in Figure 7, where there are intermediate scattering events. Also, Figure 9 shows the comparison of the unitary limit of the triplet OP with the experimental values of the compound strontium ruthenate (Miyake and Narikiyo, 1999; Nishizaki *et al.*, 1998). In this case, we recall that strontium atoms add an additional impurity level since they are part of the crystal structure. The tight binding calculation here confirms the MN original results (Miyake and Narikiyo, 1999).

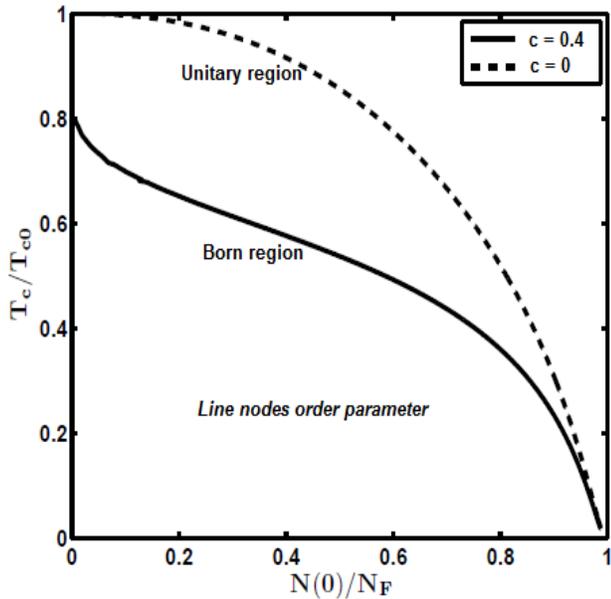

Fig. 7. The numerical calculation of the residual DoS in the case of the singlet line nodes in the Born and the unitary limits, following equations listed in Table 1.

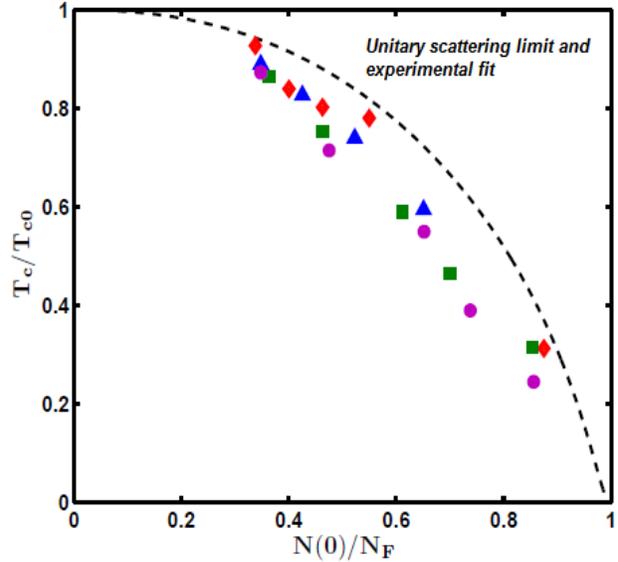

Fig. 8. The fits of the residual DoS for the unitary limit with data from La$_{2-x}$Sr$_x$CuO$_4$ (Rice and Sigrist, 1995). Different colors correspond to different hopping parameters.

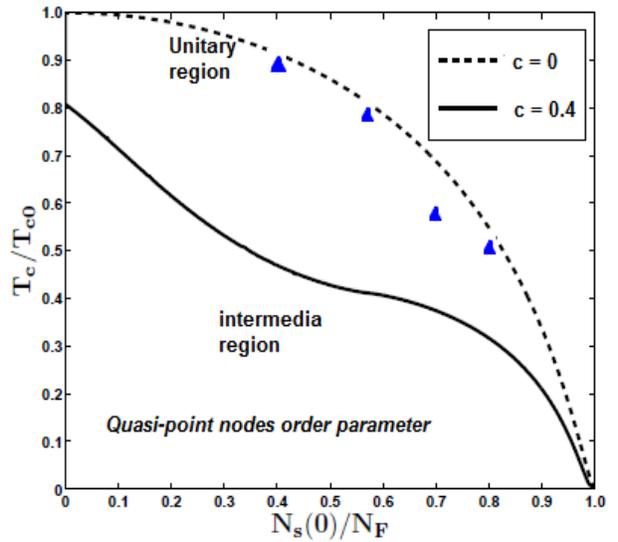

Fig. 9. The numerical calculation of the residual DoS for the triplet OP in intermedia and unitary limits. The blue color shows the experimental fits corresponding to Sr$_2$CuO$_4$ (Zhitomirsky and Rice, 2001; Walker and Contreras, 2002).

## CONCLUSION

This work was aimed at revisiting the calculation of the density of states (DoS) and the residual DoS with frequency values taken from a self-consistent calculation of the real and imaginary parts of the elastic scattering cross-section with a tight binding framework and for two order parameter models. The strontium-substituted lanthanum cuprate line nodes case (Scalapino, 1995), and strontium ruthenate symmetry breaking triplet model (Miyake and Narikiyo, 1999), where the DoS levels come from calculations in the reduced phase space when nonmagnetic pair breaking disorder destroys superconductivity. Self-consistent calculations are in general very computing demanding as stated in (Jansen *et al.*, 1991).

It was concisely reviewed main concepts and the importance of the DoS in *ab-initio* calculations for novel materials and also outlined the details of the computational approach. Also, a detailed DoS calculation of the normal state with a tight binding model was performed and interpreted in the terms of the degrees of freedom. In addition, the calculation of the superconducting DoS self-consistently performed for singlet and a triplet OP using the zero-temperature elastic cross-section for three scattering regimes. Finally, the behavior of the residual DoS was addressed for both models using the formalism following the Larkin equation (Larkin, 1965).

## RECOMMENDATIONS

It is recommended to use the tight-binding and other *ab-initio* frameworks to the study of numerical simulations as the density of states and other physical properties in doped unconventional superconductors as recently it has been done in the following work (Contreras and Osorio, 2021; Contreras *et al.*, 2022b, 2022c; Käser, 2021; Photopoulos and Frésard, 2019; Kang *et al.*, 2022; Yu *et al.*, 2023).

Concerning the approach developed in this work to obtain the imaginary part of the self-energy used to calculate the DoS, it is worthy to mention that several interesting pioneer studies about the decay of single-particle excitations in weak-coupling superconductors and the calculation of the conductivity scattering rate were conducted in the following work (Prabhu, 1978; Marsiglio and Carbotte, 1997). Referring to the unitary limit, interesting results have been obtained in the work by Domański (2011, 2016). We suggest that it would be interesting to compare the Domański results with the more recently work by Contreras *et al.* (2023) in order to see how the supercoducting zero energy gap might influence the DoS.

## ACKNOWLEDGEMENT

This research did not receive any specific grant from funding agencies in the public, commercial, or not-for-profit sectors. The authors acknowledge discussions about the presentation of the results with Dianela Osorio.

## REFERENCES

Ambegaokar, V. and Griffin, A. 1965. Theory of the thermal conductivity of superconducting alloys with paramagnetic impurities. Physical Review. 137(4A):A1151. DOI: https://doi.org/10.1103/PhysRev.137.A1151

Ashcroft, NW. and Mermin, ND. 1976. Holt, Rinehart and Winston. New York, USA. pp.826. ISBN 978-0-030-83993-1

Bardeen, J., Cooper, LN. and Schrieffer, JR. 1957. Microscopic theory of superconductivity. Physical Review. 106(1):162-164. DOI: https://doi.org/10.1103/PhysRev.106.162

Bednorz, JG. and Müller, KA. 1986. Possible high $T_c$ superconductivity in the Ba-La-Cu-O system. Zeitschrift für Physik B - Condensed Matter. 64:189-193. DOI: https://doi.org/10.1007/BF01303701

Burgos, J., Seijas, L., Contreras, P. and Almeida, R. 2017. On the geometric and magnetic properties of the monomer, dimer and trimer of NiFe$_2$O$_4$. Journal of Computational Methods in Sciences and Engineering. 17(1):19-28. DOI: https://doi.org/10.3233/JCM-160657

Contreras, P., Walker, M. and Samokhin, K. 2004. Determining the superconducting gap structure in Sr$_2$RuO$_4$ from sound attenuation studies below $T_c$. Physical Review B. 70(18):184528. DOI: https://doi.org/10.1103/PhysRevB.70.184528

Contreras, P. 2011. Electronic heat transport for a multiband superconducting gap in Sr$_2$RuO$_4$. Revista Mexicana de Física. 57(5):395-399.

Contreras, P., Burgos, J., Ochoa, E. and Uzcategui, D. 2014. A numerical calculation of the electronic specific heat for the compound Sr$_2$RuO$_4$ below its superconducting transition temperature. Revista Mexicana de Física. 60(3):184-189.

Contreras, P., Florez, J. and Almeida, R. 2016. Symmetry field breaking effects in Sr$_2$RuO$_4$. Revista Mexicana de Física. 62(5):442-449.

Contreras, P. and Moreno, J. 2019. A nonlinear minimization calculation of the renormalized frequency in dirty *d*-wave superconductors. Canadian Journal of Pure and Applied Sciences. 13(2):4765-4772.

Contreras, P. and Osorio, D. 2021. Scattering due to non-magnetic disorder in 2D anisotropic *d*-wave high $T_c$ superconductors. Engineering Physics. 5(1):1-7. DOI: https://doi.org/10.11648/j.ep.20210501.11

Contreras, P., Seijas, L. and Osorio, D. 2021. TDOS quantum mechanical visual analysis for single molecules. Canadian Journal of Pure and Applied Sciences. 15(2):5239-5245.

Contreras, PL., Osorio, D. and Ramazanov, SM. 2022[a]. Nonmagnetic tight-binding effects on the $\gamma$-sheet of $Sr_2RuO_2$. Revista Mexicana de Física. 68(2):020502. DOI: https://doi.org/10.31349/RevMexFis.68.020502

Contreras, PL., Osorio, D. and Tsuchiya, S. 2022[b]. Quasi-point versus point nodes in $Sr_2RuO_2$, the case of a flat tight binding $\gamma$ sheet. Revista Mexicana de Física. 68(6):060501. DOI: https://doi.org/10.31349/RevMexFis.68.060501

Contreras, P., Osorio, D. and Devi, A. 2022[c]. The effect of nonmagnetic disorder in the superconducting energy gap of strontium ruthenate. Physica B: Condensed Matter. 646:414330. DOI: https://doi.org/10.1016/j.physb.2022.414330

Contreras, P., Osorio, D. and Beliayev, EY. 2022[d]. Dressed behavior of the quasiparticles' lifetime in the unitary limit of two unconventional superconductors. Low Temperature Physics. 48(3):187-192. DOI: https://doi.org/10.1063/10.0009535

Contreras, P., Osorio, D. and Beliayev, EY. 2022[e]. Tight-binding superconducting phases in the unconventional compounds strontium-substituted lanthanum cuprate and strontium ruthenate. American Journal of Modern Physics. 11(2):32-38. DOI: https://doi.org/10.11648/j.ajmp.20221102.13

Contreras, P., Osorio, D. and Devi, A. 2023. Self-consistent study of the superconducting gap in the Strontium-doped Lanthanum Cuprate. International Journal of Applied Mathematics and Theoretical Physics. 9(1):1-13. DOI: https://doi.org/10.11648/j.ijamtp.20230901.11

Devi, A., Kumar, A., Singh, A. and Ahluwalia, PK. 2019. A comparative study on phonon spectrum and thermal properties of graphene, silicene and phosphorene. The AIP Conference Proceedings. 2115:030386. DOI: https://doi.org/10.1063/1.5113225

Devi, A., Kumar, A., Ahluwalia, PK. and Singh, A. 2021[a]. Novel properties of transition metal dichalcogenides monolayers and nanoribbons ($MX_2$, where M = Cr, Mo, W and X = S, Se): A spin resolved study. Materials Science and Engineering: B. 271:115237. DOI: https://doi.org/10.1016/j.mseb.2021.115237

Devi, A., Kumar, A., Kumar, T., Bharti, Adhikari, R., Ahluwalia, PK. and Singh, A. 2021[b]. Structural, electronic and magnetic properties of $Cr_mS_n$ and $Cr_mSe_n$ nanoflakes: An *ab initio* investigation. Physica E: Low-dimensional Systems and Nanostructures. 134:114825. DOI: https://doi.org/10.1016/j.physe.2021.114825

Devi, A., Kumar, N., Thakur, A., Kumar, A., Singh, A. and Ahluwalia, PK. 2022. Electronic band gap tuning and calculation of mechanical strength and deformation potential by applying uniaxial strain on $MX_2$ (M = Cr, Mo, W and X = S, Se) monolayers and nanoribbons. ACS Omega. 7(44):40054-40066. DOI: https://doi.org/10.1021/acsomega.2c04763

Domański, T. 2011. Spectroscopic Bogoliubov features near the unitary limit. Physical Review A. 84(6):023634. DOI: https://doi.org/10.1103/PhysRevA.84.023634

Domański, T. 2016. Quasiparticle states driven by a scattering on the preformed electron pairs. Condensed Matter Physics. 19(1):13701 (pages 1–11). DOI: https://doi.org/10.5488/CMP.19.13701

Edwards, SF. 1958. A new method for the evaluation of electric conductivity in metals. The Philosophical Magazine: A Journal of Theoretical Experimental and Applied Physics. 3(33):1020-1031. DOI: https://doi.org/10.1080/14786435808243244

Hussey, NE. 2002. Low-energy quasiparticles in high-$T_c$ cuprates. Advances in Physics. 51(8):1685-1771. DOI: https://doi.org/10.1080/00018730210164638

Jansen, RJE., Farid, B. and Kelly, MJ. 1991. The steady-state self-consistent solution to the nonlinear Wigner-function equation; A new approach. Physica B: Condensed Matter. 175(1-3):49-53. DOI: https://doi.org/10.1016/0921-4526(91)90688-B

Käser, S. 2021. Response functions of strongly correlated electron systems: From perturbative to many-body techniques. Ph. D. Graduate Thesis (Dr. rer. nat.). Friedrich-Alexander-Universität, Erlangen-Nürnberg, Germany. pp.233.

Kaganov, MI. and Lifshits, IM. 1989. Quasiparticles: Ideas and Principles of Quantum Solid State Physics. 2[nd] edition. Nauka, Moscow, USSR.

Kang, BL., Shi, MZ., Zhao, D., Li, SJ., Li, J., Zheng, LX., Song, DW., Nie, LP., Wu, T. and Chen, XH. 2022. NMR evidence for universal pseudogap behavior in quasi-two-dimensional FeSe-based superconductor. Chinese Physics Letters. 39(12):127401. DOI: https://doi.org/10.1088/0256-307X/39/12/127401

Kastner, MA., Birgeneau, RJ., Shirane, G. and Endoh, Y. 1998. Magnetic, transport, and optical properties of monolayer copper oxides. Reviews of Modern Physics. 70(3):897-928. DOI: https://doi.org/10.1103/RevModPhys.70.897

Landau, LD. and Lifshitz, EM. 1969. Statistical Physics. Course of Theoretical Physics. 2[nd] revised and enlarged edition. Pergamon, Pergamon Press, Oxford-New York-Toronto-Sydney-Braunschweig. Volume 5. pp.484.

Larkin, AI. 1965. Vector pairing in superconductors of small dimensions. The Soviet Physics JETP Letters. 2(5):130-132.

Lifshits, IM., Gredeskul, SA. and Pastur, LA. 1988. Introduction to the theory of disordered systems. John Wiley and Sons. pp.462.

Lu, T. and Chen, F. 2012. Multiwfn: A multifunctional wave function analyzer. Journal of Computational Chemistry. 33(5):580-592. DOI: https://doi.org/10.1002/jcc.22885

Lupien, C., MacFarlane, WA., Proust, C., Taillefer, L., Mao, ZQ. and Maeno, Y. 2001. Ultrasound attenuation in $Sr_2RuO_4$: An angle-resolved study of the superconducting gap function. Physical Review Letters. 86(26):5986-5989. DOI: https://doi.org/10.1103/PhysRevLett.86.5986

Maeno, Y., Hashimoto, H., Yoshida, K., Nishizaki, S., Fujita, T., Bednorz, JG. and Lichtenberg, F. 1994. Superconductivity in a layered perovskite without copper. Nature (London). 372:532-534. DOI: https://doi.org/10.1038/372532a0

Maeno, Y., Yoshida, K., Hashimoto, H., Nishizaki, S., Ikeda, S., Nohara, M., Fujita, T., Mackenzie, AP., Hussey, NE., Bednorz, JG. and Lichtenberg, F. 1997. Two-dimensional Fermi liquid behavior of the superconductor $Sr_2RuO_4$. Journal of the Physical Society of Japan. 66(5):1405-1408. DOI: https://doi.org/10.1143/jpsj.66.1405

Marsiglio, F. and Carbotte, JP. 1997. Quasiparticle lifetimes and the conductivity scattering rate. Australian Journal of Physics. 50(6):1011-1033. DOI. https://doi.org/10.1071/P97004

Mineev, VP. and Samokhin, KV. 1999. Introduction to Unconventional Superconductivity. Gordon and Breach Science Publishers, New York, USA. pp.200.

Miyake, K. and Narikiyo, O. 1999. Model for unconventional superconductivity of $Sr_2RuO_4$. Effect of impurity scattering on time-reversal breaking triplet pairing with a tiny gap. Physical Review Letters. 83(7):1423-1426. DOI: https://doi.org/10.1103/PhysRevLett.83.1423

Kang, BL., Shi, MZ., Zhao, D., Li, SJ., Li, J., Zheng, LX., Song, DW., Nie, LP., Wu, T. and Chen, XH. 2022. NMR evidence for universal pseudogap behavior in quasi-two-dimensional FeSe-based superconductor. Chinese Physics Letters. 39(12):127401. DOI: https://doi.org/10.1088/0256-307X/39/12/127401

Momono, N. and Ido, M. 1996. Evidence for nodes in the superconducting gap of $La_{2-x}Sr_xCuO_4$. $T^2$ dependence of electronic specific heat and impurity effects. Physica C: Superconductivity. 264(3-4):311-318. DOI: https://doi.org/10.1016/0921-4534(96)00290-0

Mulhall, D. and Moelter, MJ. 2014. Calculating and visualizing the density of states for simple quantum mechanical systems. American Journal of Physics. 82(7):665-673. DOI: https://doi.org/10.1119/1.4867489

Nishizaki, S., Maeno, Y., Farner, S., Ikeda, S. and Fujita, T. 1998. Evidence for unconventional superconductivity of $Sr_2RuO_4$ from specific-heat measurements. Journal of the Physical Society of Japan. 67(2):560-563. DOI: https://doi.org/10.1143/JPSJ.67.560

Nomura, T. 2005. Theory of transport properties in the $p$-wave superconducting state of $Sr_2RuO_4$ - a microscopic determination of the gap structure. Journal of the Physical Society of Japan. 74(6):1818-1829. DOI: https://doi.org/10.1143/jpsj.74.1818

Palczewski, AD. 2010. Angle-resolved photoemission spectroscopy (ARPES) studies of cuprate superconductors. Ph. D. Graduate Theses and Dissertations. Iowa State University, Iowa, USA. pp. 96.

Pethick, CJ. and Pines, D. 1986. Transport processes in heavy-fermion superconductors. Physical Review Letters. 57(1):118-121. DOI: https://doi.org/10.1103/PhysRevLett.57.118

Photopoulos, R. and Frésard, R. 2019. Cuprate superconductors: A 3D tight-binding model for La-based cuprate superconductors. Annalen der Physik. 531(12):1970044. DOI: https://doi.org/10.1002/andp.201970044

Prabhu, R. 1978. Decay of single-particle excitations in weak-coupling superconductors. Physica A: Statistical Mechanics and its Applications. 91(3-4):612-618, DOI: https://doi.org/10.1016/0378-4371(78)90203-0

Reif, F. 1965. Fundamentals of Statistical and Thermal Physics. McGraw Hill, New York, USA. pp.651. ISBN 0-07-051800-9.

Rice, TM. and Sigrist, M. 1995. $Sr_2RuO_4$: an electronic analogue of $^3He$? Journal of Physics: Condensed Matter. 7(47):L643-L648. DOI: https://doi.org/10.1088/0953-8984/7/47/002

Scalapino, DJ. 1995. The case for $d_{x^2-y^2}$ pairing in the cuprate superconductors. Physics Reports. 250(6):329-365. DOI: https://doi.org/10.1016/0370-1573(94)00086-I

Schachinger, E. and Carbotte, JP. 2003. Residual absorption at zero temperature in $d$-wave superconductors. Physical Review B. 67(13):134509. DOI: https://doi.org/10.1103/PhysRevB.67.134509

Sigrist, M. 2002. Ehrenfest relations for ultrasound absorption in $Sr_2RuO_4$. Progress of Theoretical Physics. 107(5):917-925. DOI: https://doi.org/10.1143/PTP.107.917

Sun, Y. and Maki, K. 1995. Transport properties of $d$-wave superconductors with impurities. Europhysics Letters. 32(4):355-359.

Tanatar, MA., Nagai, S., Mao, ZQ., Maeno, Y. and Ishiguro, T. 2001. Thermal conductivity of superconducting Sr$_2$RuO$_4$ in oriented magnetic fields. Physical Review B. 63(6):064505. DOI: https://doi.org/10.1103/PhysRevB.63.064505

Taniguchi, H., Nishimura, K., Goh, SK., Yonezawa, S. and Maeno, Y. 2015. Higher-$T_c$ superconducting phase in Sr$_2$RuO$_4$ induced by in-plane uniaxial pressure. Journal of the Physical Society of Japan. 84(1):014707. DOI: https://doi.org/10.7566/JPSJ.84.014707

Tsuei, CC. and Kirtley, JR. 2000. Pairing symmetry in cuprate superconductors. Reviews of Modern Physics. 72(4):969-1016. DOI: https://doi.org/10.1103/RevModPhys.72.969

Walker, MB., Smith, MF. and Samokhin, KV. 2001. Electron phonon interaction and ultrasonic attenuation in the ruthenate and cuprate superconductors. Physical Review B. 65(1):014517. DOI: https://doi.org/10.1103/PhysRevB.65.014517

Walker, MB. and Contreras, P. 2002. Theory of elastic properties of Sr$_2$RuO$_4$ at the superconducting transition temperature. Physical Review B. 66(21):214508. DOI: https://doi.org/10.1103/PhysRevB.66.214508

Wu, WC. and Joynt, R. 2001. Transport and the order parameter of superconducting Sr$_2$RuO$_4$. Physical Review B. 64(10):100507. DOI: https://doi.org/10.1103/PhysRevB.64.100507

Xiao, G., Streitz, FH., Gavrin, A., Du, YW. and Chien, CL. 1987. Effect of transition-metal elements on the superconductivity of Y-Ba-Cu-O. Physical Review B. 35(16):8782-8784. DOI: https://doi.org/10.1103/PhysRevB.35.8782

Yoshida, T., Hashimoto, M., Vishik, IM., Shen, ZX. and Fujimori, A. 2012. Pseudogap, superconducting gap, and Fermi arc in High-$T_c$ cuprates revealed by angle-resolved photoemission spectroscopy. Journal of the Physical Society of Japan. 81(1):011006. DOI: https://doi.org/10.1143/JPSJ.81.011006

Yu, L., Huang, S., Xing, X., Yi, X., Meng, Y., Zhou, N., Shi, Zh. and Liu, X. 2023. Critical current density, vortex pinning, and phase diagram in the NaCl-Type Superconductors InTe1–xSex (x = 0, 0.1, 0.2Chinese Physics Letters. 40(3):037403. DOI: https://doi.org/10.1088/0256-307X/40/3/037403

Zhitomirsky, ME. and Rice, TM. 2001. Interband proximity effect and nodes of superconducting gap in Sr$_2$RuO$_4$. Physical Review Letters. 87(5):057001. DOI: https://doi.org/10.1103/PhysRevLett.87.057001

Ziman, JM. 1979. Models of Disorder: The Theoretical Physics of Homogeneously Disordered Systems. 1st Edition. Cambridge University Press, Cambridge-London-New York-Melbourne, England, UK. pp.525.